\documentstyle{article}

\font\tenrm=cmr10
\font\tenit=cmti10
\font\elevenbf=cmbx10 scaled\magstep 1
\font\elevenrm=cmr10 scaled\magstep 1
\font\elevenit=cmti10 scaled\magstep 1

\renewenvironment{thebibliography}[1]
 { \tenrm
 \baselineskip=10pt
   \begin{list}{\arabic{enumi}.}
    {\usecounter{enumi} \setlength{\parsep}{0pt}
     \setlength{\itemsep}{3pt} \settowidth{\labelwidth}{#1.}
     \sloppy
    }}{\end{list}}

\begin{document}
\begin{center}{\elevenbf
STRANGENESS, CHARM AND BOTTOM IN A CHIRAL QUARK-MESON MODEL
\\}
\vglue 0.5cm
{\tenrm V.B.Kopeliovich$^*$ and  M.S.Sriram$^{**}$ \\}
{\tenit $^*$Institute for Nuclear Research of the Russian Academy of
Sciences,\\ 60th October Anniversary Prospect 7A, Moscow 117312\\
and \\$^{**}$Department of Theoretical Physics,University of Madras,\\ 
Guindy Campus, Madras 600025, India\\}
\vglue 0.6cm
\end{center}
{\rightskip=3pc
 \leftskip=3pc
 \tenrm\baselineskip=11pt
 \noindent
In this paper we investigate an $SU(3)$ extension of the chiral quark-meson 
model.
The spectra of baryons with strangeness, charm and bottom are considered
within a "rigid oscillator" version of this model.
The similarity between the quark part of the Lagrangian in the model and the 
Wess-Zumino term in the Skyrme model is noted. The binding energies of 
baryonic systems with baryon number $B=2$ and $3$ possessing strangeness or 
heavy flavor are also estimated. The results obtained are in good qualitative
agreement with those obtained previously in the chiral soliton (Skyrme) model.
 \vglue 0.2cm}
 \vglue 0.1cm
\baselineskip=12pt
\elevenrm
\section{Introduction}
The chiral soliton models, Skyrme model first of all \cite{1}, are attractive
because they are simple, elegant and allow to describe the properties of
lowest baryons with a rather good accuracy. At the same time, since the quark
degrees of freedom are excluded from the beginning, the Skyrme model is  not 
completely realistic: it is generally believed that
quarks should be explicitly present in the baryons. Consideration of more realistic 
models with explicit quark degrees of freedom included into Lagrangian 
seems to be necessary.

For the case of nonstrange baryons it was done in papers \cite{2}-\cite{6}
within the chiral quark meson model $(CQM)$,
where the mean field approximation for the quark wave functions was
an important ingredient of the model. From a theoretical point of view, the
$CQM$ models have an advantage that there is no question about the choice of
the terms in the Lagrangian responsible for the stabilization of the soliton:
the stabilization takes place due to the quark-meson interaction. Such
models are minimal in the sense that only the second order terms in chiral fields
derivatives are present in the effective Lagrangian \cite{2}-\cite{5}.
 
Here we extend such models for the consideration of baryons with strangenes,
charm and bottom, for the sector with $B=1$, first of all. These degrees of 
freedom are treated in the
same manner as in the bound state approach to heavy flavors proposed in 
\cite{7,8} and a "rigid oscillator" version of which was developed in  
\cite{9}-\cite{11}. Within this model the deviations of quark fields
and solitons into "strange" ("charm" or "bottom") directions are considered
as small ones, and a corresponding expansion of the Lagrangian is made.
The results obtained confirm the assumption concerning the smallness of
these deviations.
 
The sectors with $B=2$ and $3$ are also briefly discussed. 
Previously, the question of existence of  baryonic systems $(BS)$ with 
strangeness
different from zero was a subject of intensive studies
beginning with papers \cite{12}-\cite{15}. Some review of theoretical 
predictions, mainly for the sector with $B=2$, can be found in \cite{16}.
The question of existence of $BS$ with flavor different from $u$ and $d$  
is quite general. Charm, bottom or top quantum numbers are
also of interest, and their consideration can be performed in the framework of
chiral soliton models, in particular, the bound state approach to heavy
flavors  \cite{7}-\cite{10}. As it was shown recently, within the
rigid oscillator model the $BS$ with charm or bottom have even more
chances to be bound relative to the strong interactions than strange
baryonic systems \cite{11}. Here we present some estimates for the binding 
energies of lightest $BS$ with nontrivial flavor in the chiral 
quark-meson model and show that these estimates are in qualitative 
agreement with those obtained in \cite{10,11}.

In section $2$ we consider the $SU(3)$ extension of the chiral quark-meson
Lagrangian. In the next section, we give an explicit expression for 
the Hamiltonian of the $BS$ in the leading
order in $N_c$ in terms of the flavor (antiflavor)
excitation frequences. In section $4$ the $B=1$ sector is 
considered and hyperon
-nucleon mass differences are estimated, including the zero modes
corrections of the order of $1/N_c$. In section $5$ sectors with 
$B \geq 2$
are discussed and binding energies of some few-baryon systems are estimated.
\section{$SU(3)$ extension of the chiral quark-meson Lagrangian}
The $SU(3)$ extension of the chiral quark-meson Lagrangian density can be 
written in the following way \cite{2,3}:
$${\cal L} = i\bar{\Psi} \hat{\partial}\Psi - gF_\pi (\bar{\Psi}_L U\Psi_R +
\bar{\Psi}_R U^{\dagger}\Psi_L) - \frac{\alpha_F}{3} gF_\pi\bigl[\bar{\Psi}_L(1- \sqrt{3}
\lambda_8)U\Psi_R +\bar{\Psi}_R U^{\dagger} (1-\sqrt{3}\lambda_8) 
\Psi_L\bigr] + $$
$$+ \frac{F_\pi^2}{16}Tr l_{\mu} l^{\mu}
+\frac{F_\pi^2m_\pi^2}{16} Tr(U+U^{\dagger}-2) + 
 \frac{F_D^2m_D^2-F_\pi^2m_\pi^2}{24}Tr(1-\sqrt{3}\lambda_8)(U+U^{\dagger}-2)+$$
$$ + \frac{F_D^2-F_\pi^2}{48} Tr(1-\sqrt{3}\lambda_8)(Ul_\mu l^\mu +
l_\mu l^\mu U^{\dagger}).\eqno (1) $$

Here $\Psi$ is a triplet of quark fields, $u,d,s$, or $u,d,c$, or $u,d,b$.
$U \in SU(3)$ is a unitary matrix incorporating chiral (meson) fields, and
$l_\mu=U^{\dagger} \partial _\mu U$. In this model $F_\pi$ is fixed 
at the physical value, $F_\pi$ = $186$ Mev, and $g F_\pi$ = $500$ Mev
characterizes the effective $u,d$ quarks mass. 
The interaction between quarks is not considered explicitly, but it is present
in this mean-field description of the quarks due to the quark-meson coupling.
Here, we have included a term in the Lagrangian  which 
describes  flavor symmetry breaking $(FSB)$ in the constituent quark 
masses and in the quark-meson coupling and is proportional to a 
parameter $\alpha_F$.
The $FSB$ in the meson part of the Lagrangian is of usual form, and was
sufficient to describe the mass splittings of the octet and decuplet of baryons
\cite{17}. Here we consider first the case of flavor symmetry in decay
constants, i.e. $F_D=F_\pi$. Even for realistic values of $F_D$ the last term
in $(1)$ is small and can be omitted.

The important property of the quark part of the Lagrangian is that 
it reproduces the properties of 
the Wess-Zumino term written in the simple form by Witten \cite{18}.
First, the baryon number is given by this term and second, when the field 
$\Psi$ is rotated into "strange", or other direction, the quark
Lagrangian gives the contribution coinciding with that coming from 
the $WZ$ term in the Skyrme model.

We shall consider the collective coordinates rotation of the quark field
$\Psi$ and the meson fields incorporated into the matrix $U$, in the
spirit of the bound state approach to the description of strangeness
proposed in \cite{7}-\cite{9} and used in \cite{10,11}:
$$ \Psi(r,t) = R(t)\Psi_0(O(t)\vec{r}), \qquad U(r,t) = R(t) U_0(O(t)\vec{r}) R^{\dagger}(t), \qquad
 R(t) = A(t) S(t), \eqno (2) $$
where $\Psi_0$ is originally a two-component spinor in $(u,d)$ $SU(2)$ subgroup,
$U_0$ is the $SU(2)$ soliton embedded into $SU(3)$ in the usual way (into
left upper corner), $A(t) \in SU(2)$ describes $SU(2)$ rotations, 
$S(t) \in SU(3)$ describes 
rotations in the "strange", "charm" or "bottom" direction, and O(t) describes
rigid rotations in real space.
For definiteness we shall consider the extension of the $(u,d)$ $SU(2)$
Skyrme model in strange direction, when $D$ is the field of $K$-mesons. 
But it is clear that quite similar extension can be made in the charm or 
bottom directions, also.
$$ S(t) = exp (i {\elevenit D} (t)),  \qquad
 {\elevenit D} (t) = \sum_{a=4,...7} D_a(t) \lambda_a, \eqno (3) $$
$\lambda_a$ are Gell-Mann matrices of $(u,d,s)$, $(u,d,c)$ or $(u,d,b)$
$SU(3)$ subgroups. The $(u,d,c)$ and $(u,d,b)$ $SU(3)$ subgroups are quite analogous to
the $(u,d,s)$ one. For the $(u,d,c)$ subgroup, a simple redefiniton of 
hypercharge should be made. For the $(u,d,s)$ subgroup,
 $D_4=(K^++ K^-)/\sqrt{2}$, $D_5=i(K^+-K^-)/\sqrt{2}$, etc.
For the $(u,d,c)$ subgroup $D_4=(D^0 + \bar{D}^0)/\sqrt{2}$, etc.

Consider first the contributon due to the time dependence of the collective
rotations in the quark part of the Lagrangian:
$${\cal L}_q=\bigl[ \sum_q i\bar{\Psi} \hat{\partial}\Psi \bigr]_{collective} =
\sum_q \Psi^{\dagger}\bigl[ iS^{\dagger}\dot{S} +
{1 \over 2}S^{\dagger}\vec{\tau}
\vec{\omega}S + i (\vec{r}\vec{\Omega}\vec{\partial})\bigr] \Psi \eqno (4) $$
$\vec{\omega}$ and $\vec{\Omega}$ are the angular velocities of the isospin and usual
space rotations defined in the standard way:
$$ A^{\dagger} \dot{A} =-i \vec{\omega} \vec{\tau}/2, \qquad
\dot{O}_{in} O_{kn} = \epsilon_{ikm} \Omega_m $$
The field $D$ is small in magnitude, of order $1/\sqrt{N_c}$, 
where $N_c$ is the number of colors in $QCD$.
Therefore, an expansion of the
matrix $S$ in $D$ can be made. Collecting all the terms upto $O(1/N_c)$,
 ${\cal L}_q$ can be presented in the following form:
$${\cal L}_q\simeq  \sum_q \Psi^{\dagger}_0\bigl[{i\over 2} 
(\dot{D}^{\dagger}D- 
D^{\dagger}\dot{D})(1-{2 \over 3}D^{\dagger}D-{1 \over 2}\vec{\tau}
D^{\dagger}\vec{\tau}D) +
  {1 \over 2} (\vec{\omega}-\vec{\beta}) \vec{\tau}  
 - {1 \over 2}(\vec{\omega}\vec{\tau} D^{\dagger}D
+\vec{\omega} D^{\dagger}\vec{\tau}D )$$
$$  +{1 \over 12} D^{\dagger}D\vec{\tau}\vec{\beta} 
+i (\vec{r}\vec{\Omega} \vec{\partial} )
\bigr] \Psi_0 \eqno (5a) $$
Here 
$$\vec{\beta} = i (D^{\dagger}\vec{\tau}\dot{D}-
\dot{D}^{\dagger}\vec{\tau}D) \eqno (5b)  $$ 
is the angular velocity of rotation in the
"flavor" direction, $D$ is the doublet of heavy meson fields, kaons, $D$-
or $B$-mesons.
Expression $(5a)$ does not depend on the color of the quark directly, 
but some of the terms in $(5a)$ depend on the orientation of the
quark in isospace and on its radial wave function which need not be 
the same for all the quarks. However, for an arbitrary B, we always find 
the following term containing the factor $N_{c}$ after sumation over quarks
and integration over space:
$$ L_q = \frac{N_c B}{2}\frac{s_d^2}{d^2}\bigl[i(\dot{D}^{\dagger}D-
D^{\dagger}\dot{D}) -\vec{\omega} D^{\dagger}\vec{\tau}D \bigr], \eqno (6) $$
which is valid in any order in $d^2=2D^{\dagger}D$. 
It is assumed in $(6)$ that the quark wave functions $\Psi$ are
properly normalized. This contribution
coincides with that obtained from the Wess-Zumino term in the action
in the collective coordinates quantization procedure \cite{19,13}.

The general parametrization of $U_0$ for an $SU(2)$ soliton we use here 
is given by
$U_0 = c_f+s_f \vec{\tau}\vec{n}$ with $n_z=c_{\alpha}$, $n_x=s_{\alpha}
c_{\beta}$, $n_y=s_{\alpha}s_{\beta}$, $s_f=sinf$, $c_f=cosf$, etc.
The mass term of the Lagrangian $(1)$ can be calculated exactly, without
expansion in the field $D$ because the matrix\\
 $ S=1-iD \; sind/ d
-D^2 (1-cosd)/ d^2 $ :
$$\Delta{\cal L}_M=-\frac{F_D^2m_D^2-F_\pi^2m_\pi^2}{4} (1-c_f) s_d^2 \eqno (7)$$
The expansion of this term can be done easily up to any order in $d$.
The comparison of this expression with $\Delta L_M$ within the collective
coordinates approach allows to establish the relation $sin^2 d =sin^2 \nu$,
where $\nu$ is the angle of the $\lambda_4$ rotation, or rotation into
"strange" direction.
The so called strangeness (or flavor) content of the quark fields can be 
calculated easily, $C_s \simeq D^{\dagger}D$.
It should be remembered that in the collective coordinates method strangeness 
content of the soliton $C_s = (sin^2 \nu) /2$.
  
The time-dependent part of the second order term in the Lagrangian density 
$(1)$ due to rotations in the configuration space 
leads to the following contribution:
$${\cal L}_2 = 2 Tr [ \dot{S}\dot{S}^{\dagger}+S^{\dagger}\dot{S}
U_0^{\dagger}S^{\dagger}
\dot{S}U_0 +\dot{A}\dot{A}^{\dagger} +2 A^{\dagger}\dot{A}S\dot{S}^{\dagger}+$$
$$+S^{\dagger}A^{\dagger}\dot{A}SU_0^{\dagger}S^{\dagger}A^{\dagger}
\dot{A}SU_0+
S^{\dagger}\dot{S}U_0^{\dagger}S^{\dagger}A^{\dagger}\dot{A}SU_0+S^{\dagger}
A^{\dagger}\dot{A}SU_0^{\dagger}S^{\dagger}\dot{S}U_0] \eqno (8) $$
Making an expansion of the matrix $S$ and adding also contributions from the
usual space rotations we obtain:
$${\cal L}_2 \simeq \frac{F_\pi^2}{8} \bigr\{4(1-c_f)[\dot{D}^{\dagger}\dot{D}
(1-{2 \over 3}
D^{\dagger}D) -{2 \over 3} ( D^{\dagger} \dot{D} \dot{D}^{\dagger}D-(D^{\dagger}
\dot{D})^2 -(\dot{D}^{\dagger}D)^2 )+\vec{\omega} \vec{\beta}/2]+$$
$$ +s_f^2 [(\vec{\omega} -\vec{\beta})^2 -(\vec{\omega}\vec{n}-
\vec{\beta}\vec{n})^2]
+ (\vec{\partial}f \vec{r}\vec{\Omega})^2+s_f^2 (\vec{\partial}n_i\vec{r}
\vec{\Omega})^2+2s_f^2(\vec{\omega}\vec{n} \partial _i\vec{n})\epsilon_{ikl}
r_k\Omega_l \bigr\} \eqno (9) $$
The moments of inertia of the configuration - coefficients in the quadratic
form in angular velocities of rotation - can be extracted easily from $(9)$.

The interaction of quarks and mesons gives the contribution proportional to
the new parameter $\alpha_F$, after integration over space:
$$ L_{int} = - \alpha_F E_{qm} D^{\dagger}D(1-\frac{2}{3}D^{\dagger}D) 
\eqno (10) $$
where, according to \cite{2,3}, $E_{qm} < 0$ is the quark-meson 
interaction energy.

Equation (9) simplifies considerably for spherically symmetrical configurations
(hedgehogs) for $B=1$, as well as for $B \geq 2$ solitons described by
axially 
symmetrical configurations (see Sections 4 and 5 for details).

After some calculation, the Lagrangian  of the chiral quark-meson
 model in the lowest order in field $D$ can be written in the form below
which is similar to that of the bound state approximation to the
topological soliton model  \cite{7}-\cite{10}:
$$ L=-M_{cl,B}+4\Theta_{F,B} \dot{D}^{\dagger}\dot{D}-[\Gamma_B(m_D^2-
m_{\pi}^2)+\alpha_F E_{qm}]D^{\dagger}D -
i{N_cB \over 2}(D^{\dagger}\dot{D}-\dot{D}^{\dagger}D). \eqno(11)$$
We have ignored the effect of the difference between $F_K$ and $F_\pi$
through the last term in $(1)$, in the above expression.
We have maintained our former notation for the moment of inertia for the
rotation into "strange", "charm" or "bottom" direction $\Theta_c=
\Theta_b=\Theta_s=\Theta_{F}$ (the index $c$ means the charm quantum number,
except in $N_c$).
In the present model, this moment of inertia has a simple analytical form for 
arbitrary starting $SU(2)$ skyrmion, regardless of its symmetry properties:
$$\Theta_{F,B} = {F_\pi^2 \over 8} \int (1-c_f) d^3r. \eqno (12) $$
Note that since the Skyrme term is absent in the $CQM$ model, this formula is 
especially simple. 

The quantity $\Gamma_B$ defines the contribution
of the mass term in the Lagrangian:
$$ \Gamma_B = {F^2_\pi \over 2} \int (1-c_f) d^3r, \eqno (13) $$
so, the following relation is valid in $CQM$:
$$ \Gamma_B= 4 \Theta_{F,B} \eqno (14) $$

The term in $(11)$ proportional to $N_cB$ which comes from the quark part here,
 is responsible for the splitting
between excitation energies of strangeness and antistrangeness 
(flavor and antiflavor in general case) \cite{8}-\cite{10}.
\section{Flavor excitation frequences}
After the canonical quantization procedure the Hamiltonian of the 
system including the terms of the order of $N_c^0$, takes the form which is 
similar to that in the topological soliton models \cite{9,10}:
$$H_B=M_{cl,B} + {1 \over 4\Theta_{F,B}} \Pi^{\dagger}\Pi + \bigl[\Gamma_B 
\bar{m}^2_D+\alpha_F E_{qm}+\frac{N_c^2B^2}{16\Theta_{F,B}} \bigr] 
D^{\dagger}D +i {N_cB \over 8\Theta_{F,B}}
(D^{\dagger} \Pi- \Pi^{\dagger} D). \eqno (15) $$
$\bar{m}_D^2 = m_D^2-m_\pi^2$.
The momentum $\Pi$ is canonically conjugate to variable $D$ (see Eq.$(23)$
below).
Eq. $(15)$ describes the oscillator-type motion of the field $D$ 
in the background formed 
by the $(u,d)$ $SU(2)$ soliton. After the diagonalization which can be done
explicitely according to \cite{9,10} the normal-ordered Hamiltonian can be 
written as
$$H_B= M_{cl,B} + \omega_{F,B} a^{\dagger} a + \bar{\omega}_{F,B} b^{\dagger} b
 + O(1/N_c) \eqno (16) $$
with $a^\dagger$, $b^\dagger$ being the operators of creation of strangeness
(i.e., antikaons) and antistrangeness
(flavor and antiflavor) quantum number, $\omega_{F,B}$ and 
$\bar{\omega}_{F,B}$ being the 
frequences of flavor (antiflavor) excitation. $D$ and $\Pi$ are connected
with $a$ and $b$ in the following way \cite{9,10}:
$$ D^i= \frac{1}{\sqrt{N_cB\mu_{F,B}}}(b^i+a^{\dagger i}), \qquad
\Pi^i = \frac{\sqrt{N_cB\mu_{F,B}}}{2i}(b^i - a^{\dagger i}) \eqno (17) $$
with
$$ \mu_{F,B} =[ 1 + 16 (\bar{m}_D^2 \Gamma_B+\alpha_F E_{qm}) \Theta_{F,B}/ 
(N_cB)^2 ]^{1/2}. $$
For the lowest states the values of $D$ are small:
$$ D \sim \bigl[16\Gamma_B\Theta_{F,B}\bar{m}_D^2 + N_c^2B^2 \bigr]^{-1/4}, $$
and increase with increasing $|F|$ like $(2|F|+1)^{1/2}$
As it was noted in \cite{10}, deviations of the field $D$ from the vacuum 
decrease with increasing mass $m_D$, as well as with increasing number of 
colors $N_c$, and the method works for any $m_D$ - for charm and bottom 
quantum number also.

The excitation frequences $\omega$ and $\bar{\omega}$ are:
$$ \omega_{F,B} = \frac{N_cB}{8\Theta_{F,B}} ( \mu_{F,B} -1 ), \qquad
 \bar{\omega}_{F,B} = \frac{N_cB}{8\Theta_{F,B}} ( \mu_{F,B} +1 ) \eqno (18) $$
As it was observed in \cite{11}, the difference 
$\bar{\omega}_{F,B}-\omega_{F,B} = N_cB/(4\Theta_{F,B})$ coincides in the 
leading order in $N_c$ with that obtained in the collective coordinates 
approach \cite{15}.

To get an idea about the value of the parameter $\alpha_F$, we can write a 
relation between $\alpha_F \, gF_\pi$ and the effective quark mass:
$$ (1 + \alpha_F) g F_\pi \simeq m_F^{eff} \eqno (20) $$
Since the quark-meson interaction energy is negative - it leads to the
stabilization of the whole configuration - the term $\alpha_F E_{qm}$
makes the flavor excitation frequences smaller. The relative role of this 
effect decreases with increasing mass of the flavor, and is most important
for strange baryons.
For the $B=1$ configuration the quark-meson interaction energy, 
$E_{qm} = -1.127 \;Gev $ \cite{3}.
For strange baryons, to have constituent strange quark mass greater than that
of nonstrange quarks mass by about $\sim 0.2 \; Gev$ we should have
$\alpha_s \simeq 0.4$. Similarly, we can obtain the crude estimates, 
$\alpha_c \simeq 2.7$ and $\alpha_b \simeq 9.4$.

The $FSB$ in the flavor decay constants, i.e. the fact that $F_K/F_\pi 
\simeq 1.23$ and $F_D/F_\pi=1.7 \pm 0.2$, should be taken into account as well.
 In the Skyrme model it leads to the increase of the flavor excitation 
frequences which changes the spectra of flavored baryons in better agreement 
with data \cite{21,22}, and leads also to some changes of the binding energies 
of $BS$ 
\cite{11}. It was mainly due to the large contribution of the Skyrme term in 
the Lagrangian to the inertia $\Theta_F$. Since the Skyrme term in the $CQM$
model under consideration is absent - we obtain the relation $\Gamma_B=4 
\Theta_{F,B}$  as a result - the influence of $FSB$ in decay constants is much
less important in the chiral quark-meson model.

The terms of the order of $N_c^{-1}$ in the Hamiltonian depending on the
angular velocities of rotations in the isospin and the usual space and
describing the zero-modes contributions are not crucial but also important
for numerical estimates of baryons spectra. They will be considered in the next
Sections.
 \section{$B=1$ hedgehog and  estimates of baryon spectra}
The  $B=1$  hedgehog configuration in the chiral quark-meson model can be 
treated in same manner as the corresponding one in 
 the topological (Skyrme) model. The unit vector $\vec{n}$
characterizing the chiral meson field configuration is $\vec{n}=\vec{r}/r$,
and the spinor $\Psi_0$ has the structure
 $$ \Psi_0= \left(\begin {array}{c}
 G(r)\chi_h \\ i\vec{\sigma}\hat{r}F(r)\chi_h
\end{array}
\right)$$
where $\chi_h$ is the hedgehog spinor
       $$\chi_h={1 \over \sqrt{2}} (u\downarrow - d\uparrow). \eqno (21)$$
It can be checked that for hedgehogs the terms in $(5)$ which depend on the
orientation of the quarks in iso- and spin space, i.e. those proportional
to $\vec{\tau}$ give zero contribution into Lagrangian. Rotations in the iso-
and usual spaces are equivalent for hedgehogs, and the contribution to the
energy depends on one common moment of inertia, $\Theta_{T,B}$.

       From equations (1), (6),(9) and (10) in section 2, we obtain the
following expression for the Lagrangian including all the terms upto $O(1/N_c)$:
 $$L \simeq -M_{cl} +4 \Theta_F[\dot{D}^{\dagger}\dot{D}(1-{2 \over 3}
D^{\dagger}D) -{2 \over 3} ( D^{\dagger} \dot{D} \dot{D}^{\dagger}D-(D^{\dagger}
\dot{D})^2 -(\dot{D}^{\dagger}D)^2 )]+
 2\Theta_F(\vec{\omega} \vec{\beta})+ \frac{\Theta_T}{2} (\vec{\omega}-
          \vec{\beta})^2$$
$$- (\Gamma_B \bar{m}_F^2 +\alpha_F E_{qm})
D^{\dagger}D(1-{2 \over 3} D^{\dagger} D)+i\frac{N_cB}{2}(1-\frac{2}{3}
D^{\dagger}D)(\dot{D}^{\dagger}D-D^{\dagger}\dot{D})
-\frac{N_cB}{2}\vec{\omega}D^{\dagger}\vec{\tau}D. \eqno (22) $$
 From this expression we can find the canonical variables,
$$ \Pi =\frac{\partial L}{\partial\dot{D}^{\dagger}}=4\Theta_F \bigl[\dot{D}\bigl
(1-{2 \over 3}D^{\dagger}D\bigr)-
{2 \over 3} D^{\dagger}\dot{D} \, D+{4 \over 3}\dot{D}^{\dagger}D \, D\bigr]
+i(\Theta_T-2\Theta_F)\vec{\omega}\vec{\tau} D-i\Theta_T\vec{\beta}\vec{\tau}
D +i {N_cB \over 2}\bigl(1- {2 \over 3} D^{\dagger}D \bigr) \, D , \eqno (23)$$
 $$ \vec{I}_{bf} = \partial L / \partial \vec{\omega} =\Theta_T \vec{\omega}
+(2\Theta_F - \Theta_T)\vec{\beta} -\frac{N_cB}{2}D^{\dagger}\vec{\tau}D.
 \eqno (24a) $$
or
$$ \vec{I}_{bf}= \Theta_T \vec{\omega}+\bigl(1 
-\frac{\Theta_T}{2\Theta_F}\bigr) \vec{I}_F -\frac{N_cB \Theta_T}{4\Theta_F}
D^{\dagger}\vec{\tau}D \eqno (24b) $$
with $\vec{I}_F = (b^{\dagger}\vec{\tau}b - a\vec{\tau}a^{\dagger})/2$.

    Using the relations
$$ -i\vec{\beta}\vec{\tau}D = 2D^{\dagger}D\dot{D} - (\dot{D}^{\dagger}D
+D^{\dagger}\dot{D})D$$
and
$$\vec{\beta}^2 = 4D^{\dagger}D\dot{D}^{\dagger}\dot{D} - (\dot{D}^{\dagger}D
+D^{\dagger}\dot{D})^{2},$$
one can see that $L$, $\Pi$ and $\vec{I}_{bf}$ have essentially the same
structures as the corresponding expressions in \cite{10}. This is true for
the Hamiltonian also, and we find that  
the $\sim 1/N_c$ zero modes quantum correction to the energies of
hedgehogs in the CQM has a structure which is very similar to the correction
term in the Skyrme model and 
can be estimated according to the expression \cite{9,10}:
$$\Delta E_{1/N_c} = {1 \over 2\Theta_{T,B}}\bigl[c_{F,B} T_r(T_r+1)+
(1-c_{F,B})I(I+1) + (\bar{c}_{F,B}-c_{F,B})I_F(I_F+1) \bigr], \eqno(25) $$
where $I=I_{bf}$ is the isospin of the baryon or $BS$, $T_r$ is the quantity 
analogous to the
"right" isospin $T_r$ in the collective coordinates approach \cite{20,19},
and $\vec{T_r}=\vec{I}_{bf}-\vec{I_F}$, $\vec{I}_F={1 \over 2}(b^{\dagger}
\vec{\tau}b-a\vec{\tau}a^{\dagger})$.
$$    c_{F,B}=1-\frac{\Theta_{T,B}}{2\Theta_{F,B}\mu_{F,B}}(\mu_{F,B}-1), 
\qquad
\bar{c}_{F,B}=1-\frac{\Theta_{T,B}}{\Theta_{F,B}(\mu_{F,B})^2}(\mu_{F,B}-1).
 \eqno(26)$$ 
In the case of antiflavor excitations, we have the same formula $(25)$, with 
the substitution $ \mu \to -\mu $ in $(26)$. For example,
$$ \bar{c}_{\bar{F},B}=1 + \frac{\Theta_{T,B}}{\Theta_{F,B}\mu^2_{F,B}}
(\mu_{F,B}+1). \eqno (27) $$
According to $(9)$ the isotopic inertia 
$$\Theta_T=\frac{F_\pi^2}{6} \int s_f^2 d^3r, \eqno (28) $$ 
but it receives some contribution, about $30\%$,
also from the quark part
of the Lagrangian due to the cranking procedure described in \cite{3}.
For numerical estimates here we take the value of $\Theta_T$ obtained
in \cite{3} in the linear $\sigma$ model since the differences of 
all calculated quantities in the linear and the nonlinear versions of the
$\sigma$ model are negligible.

In the rigid oscillator model the states predicted are not identified
with definite $SU(3)$ or $SU(4)$ representations. However, it can be done,
as shown in \cite{10}.  The quantization
condition $(p+2q)/3=B$ \cite{19} for arbitrary $N_c$ is changed to 
$(p+2q)=N_cB+3n_{q\bar{q}}$, where $n_{q\bar{q}}$ is the number of
additional quark-antiquark pairs present in the quantized states. 
For example, the state with $B=1$, $|F|=1$, $I=0$ and
$n_{q\bar{q}}=0$ should belong to the octet of $(u,d,s)$, or $(u,d,c)$,
etc. $SU(3)$ group, if $N_c=3$, see also \cite{10}. 
If $\Theta_F \rightarrow \infty$, Eq. $(25)$ goes
 over into the expression obtained for axially symmetrical $BS$ in the
collective coordinate approach \cite{15}. In realistic case with 
$\Theta_T/\Theta_F \simeq 2.9$, the structure of $(25)$ is more complicated.

 We will first summarise the results for B=1 in the 'rigid oscillator'
approach to heavy flavours in CQM, without including the effect of flavour 
symmetry breaking in the quark-meson couplings (that is, $\alpha_F = 0$). 
We find that the excitation frequencies $\omega_F$ are in general higher 
than in the Skyrme model. This can be attributed to the fact that the value of
$\Gamma_B$ in the present model is higher than the value in the Skyrme model.
The mass diference $m_\Lambda - m_N$ comes out to be $284 Mev$ compared to the
experimental value of $176 Mev$. However it is to be noted that the value of
$\omega_S$ in the model in the rigid oscillator approach used here is close to
the value of $315 Mev$ obtained in a random phase approximation to $CQM$
with broken $SU(3)$ \cite {4}.
  
\vspace{2mm}
\begin{center}
\begin{tabular}{|l|l|l|l|l|l|l|l|l|l|}
\hline
 $F$  &$\omega_F$& $\omega_{Sk}$ & $\bar{\omega}_F$&$<D^{\dagger}D>$&
$\Delta M_{\Lambda_F-N}$&$\Delta M_{\Lambda_F-N}^{exp}$ 
&$\Delta M_{\Sigma_F-N}$&$\Delta M_{Z_F-N}$& $ \bar{c}_{F}$  \\
\hline
$s$&$0.326$&$0.196$&$0.69$&$0.12$&$0.28 $&$0.176$&$0.44$&$0.78$&$0.34$ \\
\hline
$c$&$1.687$&$1.18$&$2.05 $&$0.032$&$1.67$&$1.346$&$1.89$&$2.07$&$0.75$  \\
\hline
$b$&$5.098$&$3.66$&$5.461$&$0.011$&$5.09 $&$4.702$&$5.32$&$5.47$&$0.90$ \\
\hline
\end{tabular}
\end{center}
\vspace{2mm}

{\baselineskip=10pt 
\tenrm

{\bf Table} The excitation frequences for flavor $F$, $\omega_F$,
antiflavor, $\bar{\omega}_F$ and the energy differences of baryons with 
different flavors and the nucleon, in $Gev$. $\omega_{Sk}$ are the flavor
excitation frequences in the Skyrme model shown here for comparison.
 For $B=1$ soliton we use 
the values of mass $M_1=1149$ $Mev$, flavor inertia 
$\Theta_F =2.06 \; Gev^{-1}$\cite {23}
and isotopic inertia $\Theta_T=5.93 \; Gev^{-1}$ \cite{3}.
The estimate used here, $<D^{\dagger}D>=(N_cB\mu)^{-1}$, is valid for the
lowest state of oscillator with $|F|=0$, i.e. for the nucleon.}\\

\vspace{2mm}

It should be noted that the values of inertia obtained within the chiral
          quark-meson model are close to those obtained in the Skyrme
          model. E.g. the flavor inertia $\Theta_F =1.86 \, Gev^{-1}$
          in the Skyrme model with $F_{\pi}=108 Mev$ and $e=4,84$ (nucleon
          an $\Delta$-isobar masses are fitted),
          and $\Theta_F=2.03 \, Gev^{-1}$, $\Theta_T=5.55 \, Gev^{-1}$
        in the Skyrme model variant with $F_{\pi}=186 Mev$, $e=4.12$.

The $Z_F$ baryons included in the Table have $\bar{F}$ quantum number and 
are true exotic because they cannot be
made of $N_c$ valence quarks only: one $\bar{q}q$ pair is necessary for this.
These states belong to $\bar{10}$ representation of corresponding $SU(3)$
(the upper state with isospin $I=0$). The mass of the state with $S=+1$
calculated first in \cite{24} within collective coordinates approach to the
quantization of zero modes in the Skyrme model was found about $\sim 740 \,
Mev$ above the nucleon. Later this anti-strange baryon was considered
in more details in \cite{25} where the $M_Z-M_N$ mass difference was found to
be $\sim 590 \, Mev$, also within Skyrme model, but with the additional
assumption that the $N^*(1710)$ resonance is the nonstrange component of
the $\bar{10}$ of baryons. The $CQM$ model prediction for $S=+1$ baryon 
(see the Table) is in better agreement with predictions of the collective 
coordinate method \cite{24,16}.

The inclusion of $FSB$ in the quark-meson coupling improves the situation.
We take the values of the parameter $\alpha_F$ to be: $\alpha_s=0.4$,
$\alpha_c=2.7$, $\alpha_b=9.4$ which allow to obtain the effective quark
masses in the Lagrangian close to the known values. Then we obtain,
$\omega_s=0.27 \, Gev$, $\omega_c=1.58 \,Gev$, $\omega_b = 4.97 \, Gev$.
The values of the mass differences now are, in $Gev$: $\Delta M_{\Lambda -N}
=0.229(0.176)$, $\Delta M_{\Sigma -N}=0.371(0.254) $, 
$\Delta M_{\Lambda_c -N}=1.57(1.346)$, $\Delta M_{\Sigma_c-N}=1.788
(1.516)$, $\Delta M_{\Lambda_b -N}=4.968(4.702)$, $\Delta M_{\Sigma_b -N}=
5.196$, where the figures in the parantheses correspond to the experimental
values. We see that the values are now in better agreement with data. 
The relative role of the $\alpha_F$ -term decreases with increasing mass of 
the quark, as expected.
\section{Binding energy estimates for dibaryons with strangeness, charm and
bottom}
 It was shown in \cite{5,6} that, in the chiral quark-meson model there are bound 
states of solitons with $B=2$ and greater, similar to the topological
soliton models \cite{26}. Therefore, one should expect the predictions of the
dibaryons, tribaryons, etc. with different values of flavor quantum number,
$s,c$ or $b$, stable relative to strong interactions, similar to the Skyrme
model.

The structure of the toroidal configurations with $B=2$ should be 
described first.
For B=2, $\Psi_0$ has the structure
      $$ \Psi_0= \left(\begin {array}{c}
    G(\rho,z)\chi_{1,2} \\ i\vec{\sigma}\hat{r}F(\rho,z)\chi_{1,2}
\end {array}
\right)$$
where $$\chi_1={1 \over \sqrt{2(1-cos\theta cos\alpha)}}[sin\alpha\,u\downarrow
-sin\theta e^{i\phi}\,d\uparrow -(cos\alpha -cos\theta) e^{2i\phi}\,d\downarrow]$$
and  $$\chi_2={1 \over \sqrt{2(1-cos\theta cos\alpha)}}[sin\alpha\,d\uparrow
-sin\theta e^{-i\phi}\,u\downarrow +(cos\alpha -cos\theta) e^{-2i\phi} 
u\uparrow]. \eqno (29) $$
In the B=2 soliton, $N_c$ quarks are in the state $\chi_1$ and 
 $N_c$ quarks are in the state $\chi_2$. Similar considerations apply for
higher $B$. Then equations $(5)$ and $(9)$ for the Lagrangian simplify,
in particular, the terms in $(5)$ proportional $\Psi^{\dagger}\vec{\tau}\Psi$
cancel, similar to the hedgehog case.

In \cite{6,23} the following values of the binding energy of quark-meson solitons
have been obtained: $\epsilon_2=279 $, $\epsilon_3=226$ and $\epsilon_4=192 
\; Mev$ for baryon numbers $2 \,,3$ and $4$. These values can be compared with 
the values of binding energy
in the Skyrme model, $74 \,, 72$ and $14$ $Mev$, for smaller value of the
constant, $F_\pi=108 \, Mev$ \cite{26}. For $F_\pi = 186 \, Mev$ and $e=4.12$
$\epsilon_2 = 142 \, Mev$. It makes sense to give the binding energies
in units, e.g. of the mass of the $B=1$ soliton: although the symmetry
violating mass terms in the Lagrangian violate the scaling, such comparison
gives an information which does not depend strongly on the value of $F_\pi$.
In CQM, $\epsilon_2=0.24 \, M_1$, $\epsilon_3=0.20 \, M_1$ and $\epsilon_4=
0.17 \, M_1$ to be compared with $0.086 ,\; 0.083$ and $0.016 \; M_1$ in
the Skyrme model.

Let us consider here the state with $B=2$ and $|F|=2$ with the lowest value 
of isospin, $I=0$ which can belong to the $27$-plet of corresponding $SU(3)$ 
group, $(u,d,s)$ or $(u,d,c)$, etc.
For $27$-plet of dibaryons $T_r=1$, for antidecuplet $T_r=0$. 
The quantum correction due to usual space rotations, also of the order of
$1/N_c$ is exactly of the same form as obtained in \cite{15}, see
\cite{9,10}.
Since we are interested in the lowest energy states, we discuss here the
baryonic systems with the lowest allowed angular momentum, $J=0$ for $B=2$,
and $J=3/2$ for $B=3$. The latter value is due to the constraint because of
symmetry properties of the configuration. The value $J=1/2$ is allowed for
the configuration found in \cite{27}. 

For the mass of the state with $B=|F|=2$ one obtains \cite{10}:
$$ M(B=2, |27;Y=0,\, I=0>) = M_{cl} +2\omega_{F,2}+\frac{\bar{c}_{F,2}}
{\Theta_{T,2}}. \eqno (30) $$
The binding energy of this state relative to the two $\Lambda_F$-particles:
$$\epsilon (|27; Y=0,\,I=0>) =\epsilon_2+2(\omega_{F,1}-\omega_{F,2})+
\frac{3\bar{c}_{F,1}}{4\Theta_{T,1}}-\frac{\bar{c}_{F,2}}{\Theta_{T,2}}\eqno 
(31)$$
As always, we define the binding energies relative to the decay
into $B$ baryons, nucleons or flavored hyperons.

If the moments of inertia of $BS$ at small values of $B$ were proportional
to the baryon number $B$, then the values of $\mu$, excitation frequences
$\omega_F$ and coefficients $c$ would not depend on $B$ at all. In this case
the binding energy consisted only of its classical part, and some contribution
from zero modes, the difference of $\omega$'s would not contribute.
Within the $CQM$ model the moments of inertia for $B\geq 2$ have not been 
calculated, still. Therefore, we shall make a natural assumption that the
ratios of moments of inertia for different values of $B$ in the $CQM$ model
are the same as in the Skyrme model \cite{26}. For $B=2$
$\Theta_{F,2}/\Theta_{F,1}=2.038$, $\Theta_{T,2}/\Theta_{T,1}=2.053$ \cite{26}.

With  this assumption,  we obtain the following numerical 
values: $\epsilon_{\Lambda\Lambda(S=-2)}= 0.29 \;Gev$, 
$\epsilon_{\Lambda\Lambda(c=2)} =0.31 \;Gev$, $\epsilon_{\Lambda\Lambda (b=-2)}=
0.32 \, Gev$, from expression $(31)$. It should be compared with
the binding energy of the deuteron $\epsilon_{D}=351 \; Mev$ and the
binding energy of the $NN$ scattering state with $J=0$ and isospin $I=1$, 
$\epsilon_{D'}=321 \; Mev$.
After renormalization which is necessary to produce the $NN$ scattering state
on the right place, i.e. near threshold, we obtain that the strange
dibaryon with $s=-2$ is unbound but close to the threshold,
charmed as well as bottomed dibaryon are also unbound but even more near
to the $\Lambda_F\Lambda_F$-threshold. This renormalization procedure
is justified by the fact that the number of quantum effects like loop
corrections and nonzero-modes contributions have not been, but should be 
taken into account (see also discussion of Casimir energy in Conclusions).
The binding energy of the deuteron is $30 \; Mev$ instead of measured 
$2.23 \; Mev$, so $\sim 30 \; Mev$ is the uncertainty of our approach.

The dibaryons with $|F|=1$ should be considered also. The lowest states
belong to antidecuplet of corresponding $SU(3)$, $(p,q)=(0,3)$ and have
isospin $I=1/2$. They all are bound within the approach developed here,
and become close to the threshold, even unbound after the renormalization
procedure.

These results are in qualitative agreement with those obtained
in the chiral soliton models, but it should be noted that in the
Skyrme model the states with charm and bottom remain bound after such
renormalization \cite{11}.

For $\bar{35}$-plet
of tribaryons $T_r=1/2$ (for arbitrary $(p,q)$ irrep which the $BS$ belongs to
$T_r=p/2$ if $n_{q\bar{q}}=0$). $I$ and $T$ take the lowest possible values,
$0$ or $1/2$ for $|F|$=1, and $1/2, \; 0$ for $|F|=2$. The binding energies
are of the same order of magnitude as for the $B=2$ case if we make similar
assumption concerning the behaviour of moments of inertia. 
But after renormalization the flavored states become unbound, 
although very close to thresholds.
\section{Conclusions}
We found that, as far as we are concerned with the spectra of baryons,
there is no difference of principle between topological (Skyrme) soliton
models and chiral quark meson model \cite{2,3}.
The $CQM$ model is more realistic, but, as is usual for more
realistic models, it contains an additional parameter 
which defines the flavor symmetry
breaking in the part of the Lagrangian describing the quark-meson interaction.
When this parameter is omitted, the flavor excitation frequences are too large
in comparison with the data and with the topological Skyrme model also.
Reasonable values of this parameter make the excitation frequencies
smaller, in better agreement with data.

We have estimated the spectra of baryons with flavor different from $u,d$
in the simplest $SU(3)$ extension of the chiral quark-meson model
proposed in \cite{2,3}.
One can note that the approach developed here - the rigid
oscillator version of the $CQM$ - works even better for
$c$ and $b$ flavor in comparison with strangeness.

There are predictions of the baryonic systems with $B=2,\,3...$ and flavors
$s,\,c, \, b$ similar to that in topological soliton (Skyrme) models 
\cite{15,10,11}.
In the $CQM$ model, due to the absence of the Skyrme term in the Lagrangian,
the attraction of heavy flavors by $(u,d)$ solitons is, after all the 
renormalization procedures, somewhat weaker than in topological models.
Similar predictions can also be made  for  systems with top-number.
However, because of the large width of the $t$-quark, the spectroscopy of
the baryonic systems as well as hadrons containing the $t$-quark will not
be available, most probably.

The apparent drawback of the approach exploited in the present paper is
that the motion of the system into the "strange", "charm" or "bottom" 
direction is considered independently from other motions. Consideration of
the $BS$ with "mixed" flavors is  possible in principle, but it demands a  
more complicated treatment, technically.

There is some difference between the rigid oscillator variant of the
$CQM$ we considered here and the collective coordinates approach to soliton
models widely exploited previously.
In the collective coordinates approach to the  zero modes of solitons with 
a rigid or a soft rotator variant of the model, the 
masses of baryons are usually considerably greater than in the bound state 
approach, when the Casimir energies are not taken into account
 \cite{28,29}. One of the sources of this difference is the presence
of a term of order $N_c/ \Theta_F$ in the zero-modes contribution to 
the rotation energy, which is absent in the bound state  
model. It was shown recently by Walliser for the $B=1$ sector within the
$SU(3)$ symmetrical $(m_K=m_\pi)$ variant of the Skyrme model \cite{29}
that, 
this large contribution is cancelled almost completely by the kaonic 1-loop 
correction to the  
zero-point Casimir energy which is of the same order of magnitude, $N_c^0$ 
\cite{29}. This correction has been calculated recently also within the bound
state approach to the Skyrme model \cite{30}.
The consideration of loop corrections to the energies of quantized states
is necessary also in the hybrid models similar to $CQM$.

Recently it was shown within the Skyrme model \cite{31} that, one should expect 
the existence of strange baryonic systems close to the strong decay
threshold,
for baryon numbers up to $17$. They are obtained by means of quantization
of bound $SU(2)$ skyrmions found previously in \cite{27,32}. 
The charmed baryonic systems with $B=3, \, 4$ were considered in \cite{33}
within a potential approach. The $B=3$ systems were found to be
very near the threshold and the $B=4$ system was found to be stable 
relative to the strong
decay, with a binding energy of $\sim 10 \, Mev$.

Experimental searches for the baryonic systems with flavor different from
$u$ and $d$ could shed more light on the dynamics of heavy flavors in
few-baryon systems.
The threshold for the charm production on a free nucleon is about $12 Gev$,
and for double charm it is  $\sim 25.2 \, Gev$. For bottom, the threshold on
nucleon is  $\sim 70$ $Gev$. However, for nuclei as  targets the
thresholds are much lower due to two-step processes with mesons in
intermediate states and due to normal Fermi-motion of nucleons inside the 
target nucleus (see, e.g. \cite{34}). Therefore, the production of
baryons or baryonic systems with charm and bottom will be possible on 
accelerators with energy
of several tens of $Gev$.\\

{\elevenbf Acknowledgements}\\
The authors are very indebted to J.Segar for help in numerical computations
and to H.Walliser for useful discussions.

One of the authors (MSS) thanks the Russian Academy of Sciences and The 
Indian National Science Academy for financial support which enabled him to work
 at INR, Moscow.\\

{\elevenbf\noindent References}
\vglue 0.2cm

\end{document}